\documentclass[12pt]{article}
\usepackage{lmodern}

\usepackage{xr}
\usepackage[margin=1in]{geometry}
\usepackage[utf8]{inputenc}

\usepackage{graphicx} 
\usepackage[square,numbers]{natbib}
\usepackage{authblk}
\usepackage{amsmath} 
\usepackage{newtxtext} 
\usepackage{hyperref}

\usepackage[table,x11names]{xcolor}
\usepackage{authblk}
\usepackage{booktabs}
\usepackage{threeparttable}
\usepackage{siunitx}
\usepackage{dcolumn}
\usepackage{rotating}

\usepackage{caption}
\usepackage{setspace}
\usepackage[font=small,labelfont=bf]{caption}
\usepackage{subcaption}

\usepackage{tabularx}
\usepackage{array}

\usepackage{float} 

\title{Knowledge Assemblies}

\title{\textbf{\LARGE Unraveling Human Capital Complexity: Economic Complexity Analysis of Occupations and Skills}\\
\vspace{0.3cm}  
}

\vspace{1cm}
\author[1]{Soohyoung Lee}
\author[2,3*]{Dawoon Jeong}
\author[1]{Jeong-Dong Lee}
\affil[1]{Technology Management, Economics and Policy, Seoul National University, Seoul}
\affil[2]{Kellogg School of Management, Northwestern University, Evanston, IL}
\affil[3]{Northwestern Institute on Complex Systems, Evanston, IL}
\affil[*]{Correspondence can be sent to dawoon.jung@kellogg.northwestern.edu.} 

\date{\today}

\begin{document}

\maketitle
\section*{Abstract}
This study investigates the structural embeddedness of skills in the division of labor. Drawing on O*NET data covering 120 skills across 872 U.S. occupations, we identify three skill communities: general, cognitive, and physical skills. Compressing the connectivity in the occupation--skill network through the Method of Reflection, we derive the Occupational Complexity Index (OCI) and the Skill Complexity Index (SCI). They unpack the structure of the occupation--skill network that general skills are embedded at the core, while cognitive and physical skills diverge in opposite directions. We further assess each skill’s contribution to the network’s modular and nested structure, finding that cognitive and physical skills contribute equally to specialization but differ in their interactions with general skills. Regression analysis reveals that general skills significantly moderate the wage effects of specialized skills, amplifying the returns to cognitive skills and mitigating the penalties of physical skills. These findings underscore the central function of general skills in transforming individual competencies into labor market value. Reskilling policies aimed at investing in human capital should consider general skills, which are intangible yet play a foundational role in the labor market.

\vspace{\baselineskip}
\noindent\textbf{Keywords:} Skill; Occupation; Network; Complexity; General Skill

\newpage

\maketitle
\section*{Introduction}

The division of labor has long been recognized as a key driver of modern economic growth. In the labor market, individuals can earn returns by performing occupation-specific tasks using their own skill sets. While possessing a broader range of skills may lead to greater rewards, individuals face inherent limits in their capacity to acquire multiple skills. To overcome this constraint and achieve outcomes more, workers in the labor market have historically generated synergies by exchanging their specialized competencies \cite{neffke2019value}. In this respect, beyond specialization itself, task trading enabled by social skills plays a crucial---yet often overlooked---role in the division of labor \cite{deming2017growing}.

\textit{"No occupation is an island, entire of itself."}\footnote{This phrase reinterprets John Donne's famous line to emphasize the relational nature of occupations in the labor system.} Consider the case of \textit{Lawyers}. Their value arises not only from proficiencies in the knowledge of \textit{Law and Government}, but also from their ability to apply general skills—such as \textit{coordination} or \textit{active listening}—that are transferable across seemingly unrelated occupations, including \textit{Financial Managers}, \textit{Travel Guides}, and \textit{Printing Press Operators}. In contrast, occupations such as \textit{Rail Car Repairers} depend heavily on specialized physical skills that are structurally isolated. This isolation prevents them from increasing their wages through synergies with workers in other occupations.

Recent studies have employed tools from network science to analyze the dynamics in the labor market through occupations and their skill sets. These approaches have revealed important labor market properties—such as skill polarization \cite{alabdulkareem2018unpacking}, resilience \cite{moro2021universal}, occupational relatedness \cite{farinha2019related}, constraints on labor mobility \cite{frank2024network,limfrank2023location}, and automation-induced dynamics \cite{dworkin2019network,lee2024automation,christenko2022automation,frank2025ai,frank2018small}. However, most studies focus narrowly on specialized skills, neglecting the role of general skills---those broadly required across occupations but less visible in wage metrics \cite{eggenberger2018specificity,gathmann2010general,herve2023specialists,nedelkoska2019skill}. On the other hand, while prior work such as Hosseinioun et al. \cite{hosseinioun2025skill} emphasizes the foundational role of general skills in supporting specialized ones, their structural position within the labor market remains insufficiently understood.

To fill this gap, we examine the structural embeddedness of skills in the division of labor by constructing and analyzing an occupation–skill bipartite network using O*NET data covering 872 occupations and 120 knowledge, skills, and abilities (hereafter, skills). We first binarize the importance scores of skills to identify whether each occupation requires a given skill. Based on co-occurrence proximities \cite{alabdulkareem2018unpacking,hidalgo2007product}, we detect three robust skill communities---general, cognitive, and physical---using the Louvain algorithm (see Fig.~\ref{fig:Figure 1}). General skills, such as social interaction and basic literacy, are highly ubiquitous across occupations. In contrast, specialized skills are required in only a limited set of occupations, and they are clustered into two distinct categories: cognitive (e.g., analysis, science, reasoning) and physical (e.g., manual dexterity, stamina, mechanical ability).

We then apply the Method of Reflections (MoR) \cite{hidalgo2009building} to this bipartite network to derive two complexity measures: the Occupational Complexity Index (OCI) and the Skill Complexity Index (SCI) (see Fig.~\ref{fig:Figure 2}). These indices capture how deeply an occupation or skill is embedded within the broader network structure, reflecting not only their direct connections but also higher-order interdependencies. Through iterative compression of the network using MoR, we find that high-OCI occupations are concentrated in domains such as management, finance, and IT, while low-OCI jobs dominate in production and transportation sectors. Similarly, high-SCI skills---such as mathematics, critical thinking, and systems analysis---are distinguished from low-SCI skills like trunk strength or operation monitoring.

Beyond offering new complexity indicators, the MoR-based analysis reveals deeper structural insights. The complexity indices generated by the MoR correspond to the eigenvector associated with the second-largest eigenvalue of a matrix. It captures both the similarity among nodes and the overall structure of the network, in a manner analogous to spectral clustering \cite{mealy2019interpreting}.  Plotting the occupation–skill network with OCI and SCI reveals its dual structural properties: \textit{modularity}, which reflects distinct specialization domains, and \textit{nestedness}, in which general skills link both low- and high-skilled occupations. High-OCI occupations rarely depend on low-SCI skills, and vice versa, forming voids in the matrix that delineate cognitive vs. physical specialization (see Fig.~\ref{fig:Figure 3}). At the core of this structure lie general skills, which bridge across these modules and reinforce the nested connectivity of the system.

To quantify the structural role of skills, we compute each skill’s contribution to the modularity and nestedness of the network. Cognitive and physical skills contribute equally to modular specialization. However, they diverge in nestedness: cognitive skills align more closely with general skills, whereas physical skills remain relatively isolated. These findings indicate that a skill’s value depends not only on its type, but also on how well it interacts with general skills shared across the labor system.

Finally, we examine the implications of these structural patterns for labor market returns. Regression analysis reveals that general skills play a pivotal moderating role: they amplify the wage premium of cognitive skills and mitigate the wage penalties of physical ones (see Fig.~\ref{fig:Figure 4}). This suggests that the returns to specialization are contingent on general skills that facilitate coordination, communication, and adaptability. In essence, general skills function as structural enablers, allowing workers to convert specialized competencies into tangible economic value \cite{deming2017growing}.

Taken together, our findings highlight that human capital is not isolated individual attributes, but is structurally embedded in the division of labor through interdependent skills. General skills are not merely transferable assets; they are central nodes in the labor market’s architecture, enabling the integration of specialized competencies across occupations. By facilitating coordination, adaptability, and knowledge exchange, general skills increase the value and resilience of occupational systems as a whole.

These insights offer clear implications for policy. Workforce development and reskilling programs should not only target job-specific technical skills, but also prioritize general skills that support cross-functional learning and mobility. In a labor market increasingly shaped by automation and shifting skill demands, investing in foundational, structurally embedded capabilities is essential for ensuring long-term adaptability and inclusive economic opportunity.

\maketitle
\section*{Results}

\subsection*{Identifying General Skills}

\begin{figure}[H]
    \centering
    \includegraphics[width=\textwidth]{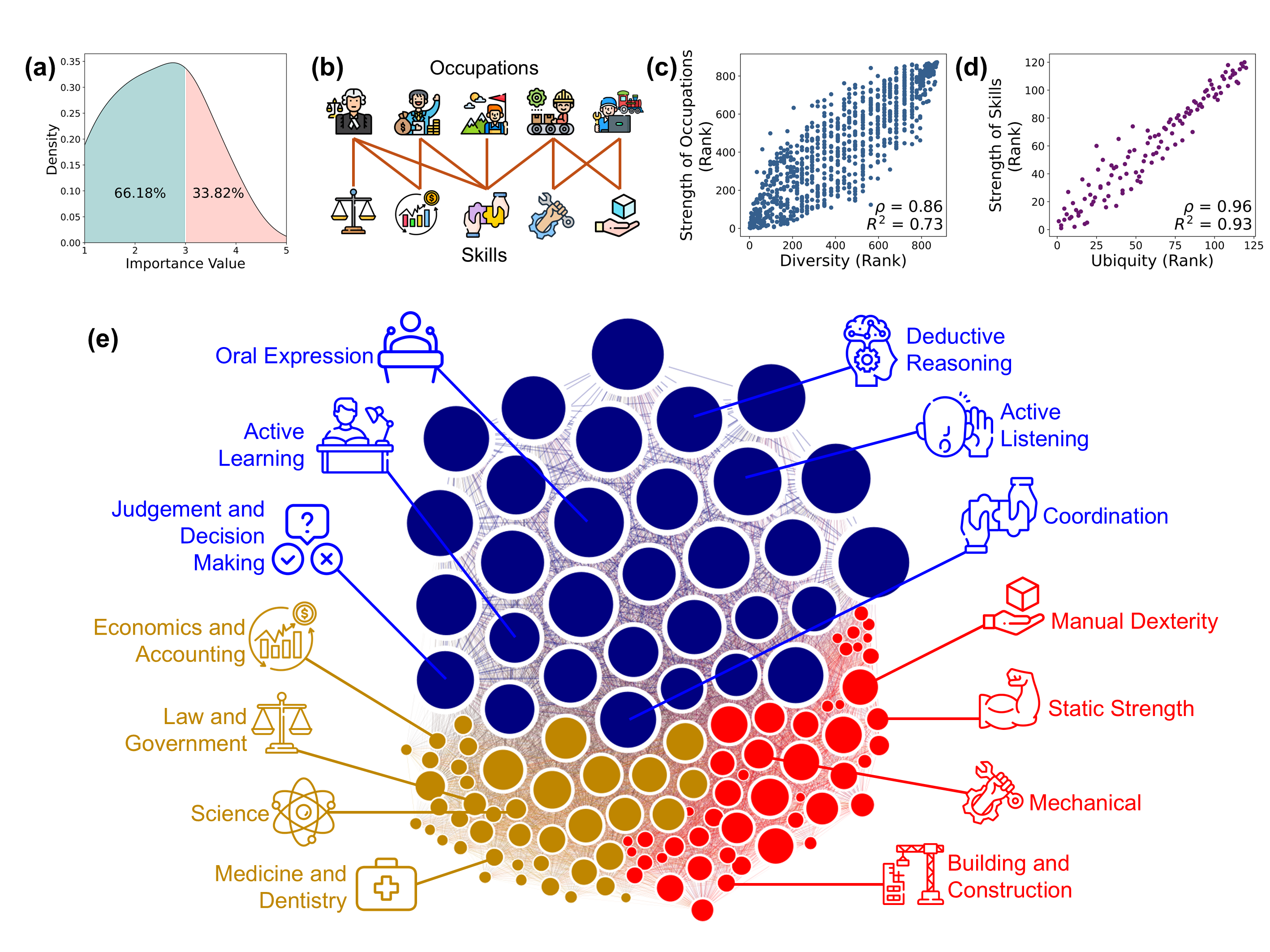}
    \caption{\textbf{General and Specialized Skills on Skill Network} \textbf{(a)} Distribution of Skill Importance Values. The KDE plot is generated with a bandwidth adjustment factor of 5. The values in the plot represent the proportions of the number of importance values less than 3 and those greater than or equal to 3, relative to the total count. Figure~\ref{fig:Figure S1} presents the distribution of skill importance values by individual 2-digit Major Classification. \textbf{(b)} Schematic representation of occupation--skill bipartite network. \textbf{(c)} The relationship between rank of diversity from binary network (the number of skills that an occupation requires) and rank of occupations' strength (the sum of importance values of skills which an occupation requires). \textbf{(d)} The relationship between rank of ubiquity from binary network (the number of occupations which requires given skill) and rank of skills' strength (the sum of the importance values of that skill required across all occupations). \textbf{(e)} The Skill Network is partitioned into three communities, consisting of 31 general skills, 41 cognitive skills, and 48 physical skills. Node sizes reflect skill ubiquity, measured by the number of occupations that require each skill.}
    \label{fig:Figure 1}
\end{figure}

Specialized (or specific) skills, which are directly associated with a worker’s competency, and general skills, which enhance the resilience of the labor market, are traditionally treated as distinct categories \cite{eggenberger2018specificity,gathmann2010general,herve2023specialists,nedelkoska2019skill}. While general skills have remained relatively overlooked, prior network-based studies of the labor market have often emphasized over-expressed skills within each occupation by normalizing skill importance data \cite{alabdulkareem2018unpacking}. Such normalization is an appropriate approach when identifying the skills a worker needs to be competitive in the labor market. However, this approach tends to render general skills---which contribute widely across the occupations---more implicit by emphasizing the only distinctive features of each occupation.

We draw attention to the fact that the skill importance data in O*NET provides scaled values based on respondents’ ratings of each skill’s importance for a given occupation on a 1--5 Likert scale. Figure~\ref{fig:Figure 1} (a) shows that the distribution of the data is centered around 3. The distribution across Standard Occupational Classification (SOC) 2-digit groups also aligns with the overall distribution (Fig.~\ref{fig:Figure S1}). Based on this observation, we set 3 as the threshold: values greater than or equal to 3 are recoded as 1, and those below 3 as 0. This binarization clarifies the skill requirements of each occupation. The resulting occupation--skill binary matrix, whose schematic representation as bipartite network is shown in Figure~\ref{fig:Figure 1} (b), consists of 872 occupations and 120 skills, after removing one occupation (O*NET-SOC Code 41-9012.00). 

Figures~\ref{fig:Figure 1} (c) and (d) demonstrate that our binary matrix preserves the information on the connectivity between occupations and skills provided by the original data (Fig.~\ref{fig:Figure S2}). The rank of diversity of occupations (the degree of occupation nodes in the network) and the strength of occupation nodes (sum of skill importance values within occupation in the network) are highly associated. Also, the rank correlation between the ubiquity of skills (the degree of skill nodes) and the strength of skills nodes (sum of skill importance values across occupations) is nearly one with almost perfect explanatory power ($R^2 = 0.93$). Such correlation and explanatory power is not captured under an approach that focuses on over-expressed skills (Fig.~\ref{fig:Figure S2}). This suggests that an approach focusing solely on specialized skills may miss important information about the labor market. In contrast, the binarization method adopted in this study can capture key aspects of the labor market.

Using the binary matrix, we examine the landscape of skills. Figure~\ref{fig:Figure 1} (e) presents the skill network constructed by computing pairwise proximity between skills (see Data and Methods). Applying the Louvain algorithm---an optimization algorithm commonly used for community detection---to the skill network, three stable communities are extracted: one comprising Basic and Social skills such as \textit{Oral Expression}, \textit{Active Listening}, and \textit{Coordination}; another including Knowledge and Cognitive Abilities such as \textit{Economics and Accounting}, \textit{Science}, and \textit{Medicine and Dentistry}; and a third dominated by Physical, Sensory, and Psychomotor Abilities such as \textit{Manual Dexterity}, \textit{Static Strength}, and \textit{Mechanical}. This community structure amplifies the findings of Hosseinioun et al. \cite{hosseinioun2025skill}, who investigate the foundational role of general skills in the directed skill network. 

Drawing on the categorization used in prior studies \cite{alabdulkareem2018unpacking,hosseinioun2025skill}, we label these communities as general, cognitive, and physical skills, respectively. Among these, the general skills community exhibits a considerable difference in ubiquity relative to the other two, indicating that general skills are required by a wide range of occupations, whereas specialized skills are needed exclusively for specific occupations. While highly ubiquitous general skills form a single undifferentiated cluster, specialized skills are divided into two distinct clusters based on their proximity. The distinct communities of the two specialized skills suggest that they are required by different occupational groups in the labor market.

These findings provide empirical evidence that general skills---despite often being overlooked---form a distinct cluster within the skill network, shared across a broad range of occupations. In contrast, specialized skills are differentiated along cognitive and physical dimensions. The classification derived from network-based community detection clearly reveals the manifestation of general skills in the labor market, aligning with the distinction between general and specialized skills in human capital theory. These results imply that focusing solely on specialized skills may overlook important information about the mechanisms in the  division of labor.

\subsection*{Structural Embeddedness of Skills in the Labor Market}

\begin{figure}[H]
    \centering
    \includegraphics[width=\textwidth]{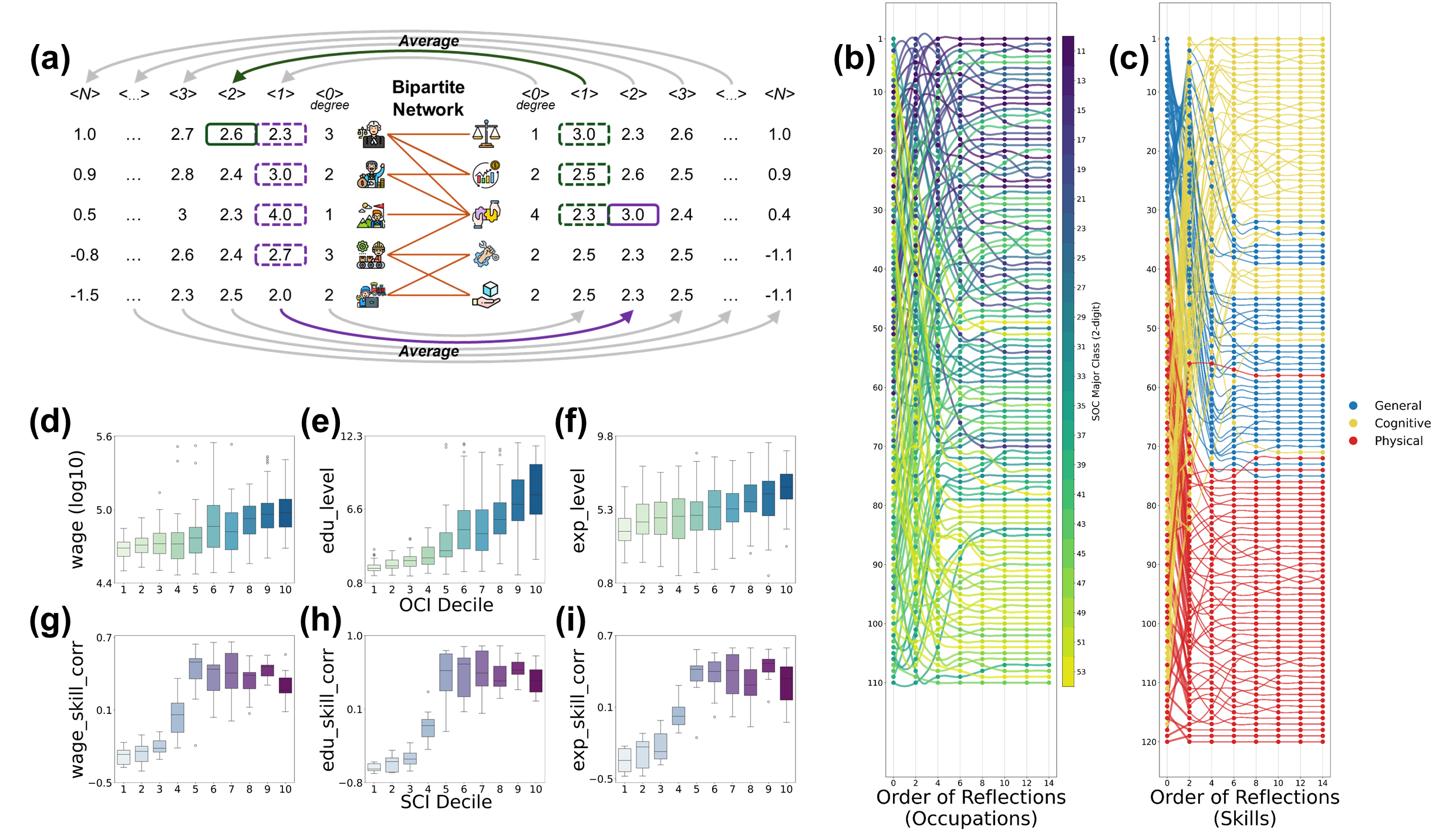}
    \caption{\textbf{Method of Reflections and Complexity of Occupations and Skills} \textbf{(a)} Method of Reflections on occupation--skill network. The standardized values are presented at N-orders. Ranking of \textbf{(b)} Occupations and \textbf{(c)} Skills by Method of Reflections orders. The values of 8-digit O*NET-SOC occupation at each order is aggregated by each 4-digit occupation. The results of even-order variables are presented for clarity. \textbf{(d)}--\textbf{(i)} The distributions of wage, required level of education and related work experience of occupations and skills by OCI and SCI decile. The skill-level variables in \textbf{(g)}--\textbf{(i)} are the correlation coefficients between the importance of each skill across occupations and the values of occupation-level variables. For the categorical variables representing education and related work experience, we follow the method of Hosseinioun et al. \cite{hosseinioun2025skill} to convert them into a continuous scale.}
    \label{fig:Figure 2}
\end{figure}

This section explores the structural embeddedness of skills within the labor market by utilizing the Method of Reflections (MoR), a conventional algorithm of economic complexity \cite{hidalgo2009building}. Given that occupations are bundles of skill requirements \cite{lee2024automation,nedelkoska2019skill,stephany2024price,acemoglu2011skills,frank2019toward}, MoR is well-suited for compressing the connectivity between occupations and skills in the bipartite network (see Data and Methods for details). The resulting one-dimensional vector is comparable to the outcomes of spectral clustering, which reflect not only the similarity between nodes but also the overall structure of the network \cite{mealy2019interpreting}.

Figure~\ref{fig:Figure 2} (a) illustrates how the MoR compresses the connectivity of occupations and skills from their own degree on the bipartite network. A node’s value is updated by averaging the values of its neighboring nodes from the previous order. For example, the second-order value of \textit{Lawyers} is the average of the first-order values of its required skills (green boxes). For the \textit{Coordination} skill at second-order has the averaged value of the occupations that require it (purple boxes). As the order increases, the value of a focal occupation (skill) increasingly condenses information not only about the skills (occupations) directly connected to it, but also about the occupations (skills) that are indirectly connected through those directly connected skills (occupations). Consequently, nodes with similar connectivity exhibit similar values at higher orders, while dissimilar nodes diverge significantly.

As shown in Figure~\ref{fig:Figure 2} (b) and (c), the value of each node converges at the more higher order. We define the standardized higher-order values of occupations and skills as the Occupational Complexity Index (OCI) and the Skill Complexity Index (SCI), respectively. The rank of OCI aligns with the Standard Occupational Classification (SOC), which generally reflects the skilled levels of occupations (Fig.~\ref{fig:Figure 2} (b)). This suggests that the OCI captures the hierarchical sophistication of occupational skill requirements by reducing the occupation--skill network to a one-dimensional vector. Figures~\ref{fig:Figure 2} (d)--(f) show the distributions of wages, educational requirements, and required work experience across OCI deciles, confirming that the OCI mirrors the occupational skilled levels.

In the same way, the SCI compresses the occupation--skill network into a one-dimensional representation with respect to skills. Its greater contribution lies in enabling the quantification of skills---particularly those that are intangible and thereby challenging to measure \cite{deming2017growing,frank2019toward}---through a vector that reflects the sophistication of skills in a hierarchical manner. Figure~\ref{fig:Figure 2} (c) shows that cognitive skills---such as \textit{Psychology}, \textit{Management of Financial Resources}, and \textit{Biology}---are intermingled with physical skills like \textit{Repairing}, \textit{Transportation}, and \textit{Manual Dexterity} in the lower ranks of the initial order. MoR iteratively compresses the information on which each skill is connected to occupations, resulting in their values diverging in opposite directions. At higher orders, cognitive skills are clustered toward the top and physical skills toward the bottom, reflecting the conventional hierarchy of skill sophistication. Figure~\ref{fig:Figure 2} (g)--(i) demonstrate that the skill hierarchy quantified by the MoR aligns with wages, education, and experience---traditional proxies for labor. This result indicates that the sophistication of skills can be quantified based on the types of occupational skill bundles in which they are embedded.

\begin{figure}[H]
    \centering
    \includegraphics[width=\textwidth]{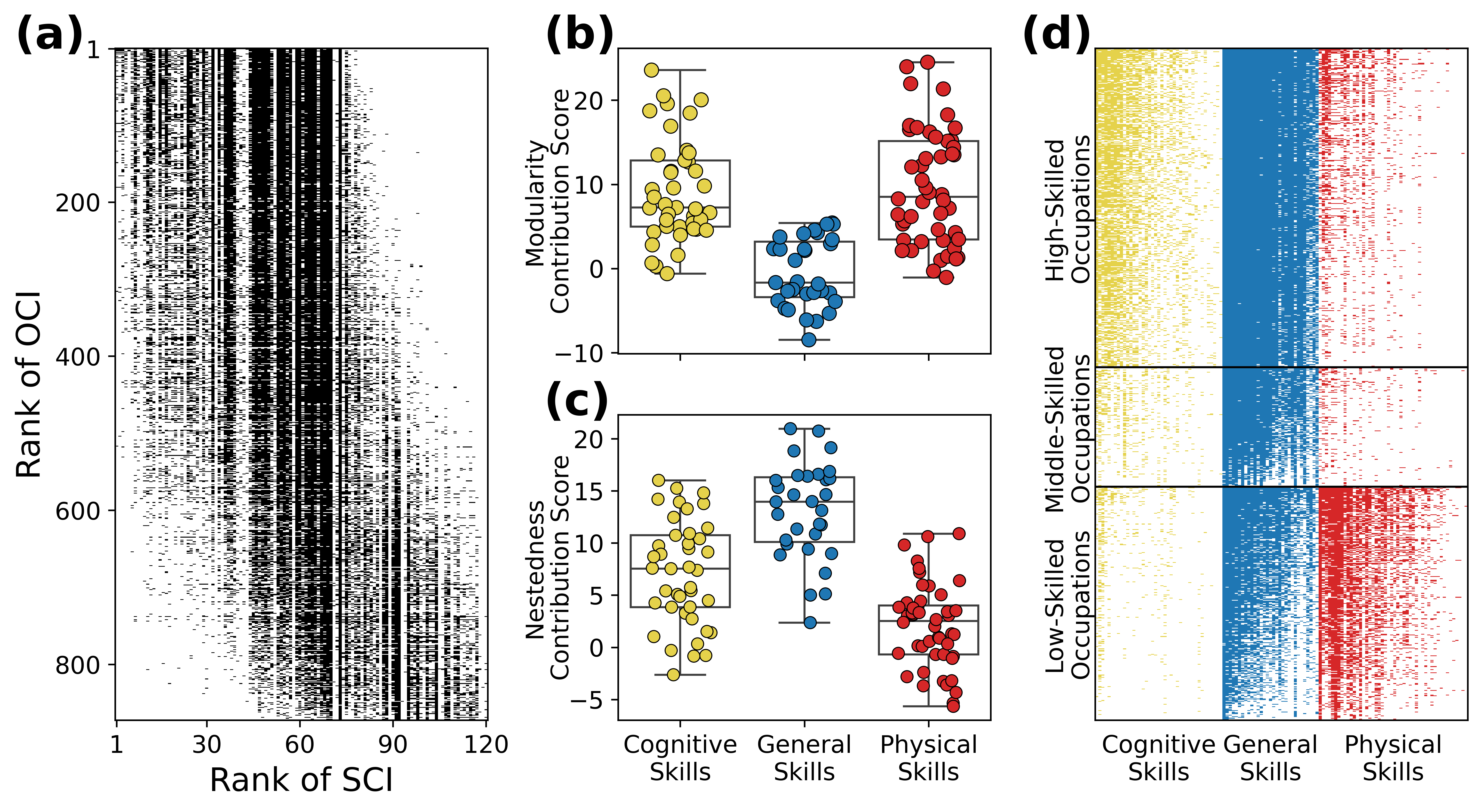}
    \caption{\textbf{Structure of occupation--skill Network} \textbf{(a)} illustrates the occupation--skill network as heatmaps. Rows and columns are ordered by the rankings of the Occupational Complexity Index (OCI) and the Skill Complexity Index (SCI), respectively. \textbf{(b)} and \textbf{(c)} show the contribution scores of each skill to the network’s modularity and nestedness, respectively. In \textbf{(d)}, occupations and skills are grouped by their communities identified from the unipartite projections of the network (Fig.~\ref{fig:Figure 1} (e) and Fig.~\ref{fig:Figure S5} (a)). A dot in heatmaps of \textbf{(a)} and \textbf{(d)} indicates the connection between occupations and skills.}
    \label{fig:Figure 3}
\end{figure}

Figure~\ref{fig:Figure 3} (a) unpacks the structure of the occupation--skill network compressed into the complexity indices. In the network illustrated as the heatmap, occupations and skills are arranged according to the rankings of OCI and SCI, respectively. Although the MoR do not produce distinct clusters (Fig.~\ref{fig:Figure S3}), the resulting heatmap can be distinguished to three regions: top--left corner, bottom--right corner, and the core center. This structure demonstrates the embeddedness of skills---two specialized skills situated at opposite corners, and general skills are centrally embedded in the labor market structure. Such a pattern is not captured by approach that focus on specialized skills (Fig.~\ref{fig:Figure S4}). 

The occupation--skill network structure, unpacked through MoR, reflects a division of labor shaped by specialization and the exchange of competencies. Skills with high-SCI values are typically connected to occupations with high-OCI values, whereas those with low-SCI values are linked to occupations with low-OCI. This pattern reflects specialization in the labor market: cognitive skills are concentrated in high-skilled occupations, while physical skills are largely confined to low-skilled ones. More importantly, the occupation--skill network highlights that general skills are embedded at the core of the labor market serve as a bridge connecting the two specialized skills. This suggests that the competencies of workers specialized in different occupations can be exchanged through general skills, thereby generating synergies that help overcome the limits of individual human capacity \cite{neffke2019value}.

Furthermore, the embeddedness of skills illustrates that two structural properties coexist within the occupation--skill network---modularity and nestedness \cite{hosseinioun2025skill}. Modularity captures a structure where interactions are dense within exclusive modules but sparse between them. Nestedness describes a systemic pattern of interactions between generalist and specialist species, thereby enhancing the survivability of specialists whose interaction with generalists is fluent in the ecosystems \cite{mariani2019physicsreport}. In Figure~\ref{fig:Figure 3} (a), the two types of specialized skills form distinct modules at opposite extreme corners. Notably, the modules independently give rise to its own nested structure with highly ubiquitous general skills, indicating the presence of two distinct patterns of nestedness within the network.

We examine the extent to which individual skills contribute to the structural properties of the network (see Data and Methods). Figure~\ref{fig:Figure 3} (b) shows that the two specialized skills contribute equally to the modularity of the network, whereas the general skills do not. This score confirms that cognitive and physical skills shape their own modules in the network and implies that they do not differ in the degree of specialization. In contrast, Figure~\ref{fig:Figure 3} (c), which presents the contribution of each skill to nestedness, reveals a clear gradient: general skills contribute the most, followed by cognitive and physical skills. This suggests that general skills serve as the backbone of interaction in the labor market, and among the two specialized skill types, cognitive skills interact more closely with general skills than physical skills do.

A more distinguishable structure of the labor market is revealed by dividing the occupation--skill network into nine blocks based on the communities of skills and occupations. Occupations are also classified into three communities, analogous to the skill communities. We label these communities as high-, middle-, and low-skilled occupations according to the distribution of their OCI values (Fig.~\ref{fig:Figure S5}). In Figure~\ref{fig:Figure 3} (d), the cognitive--high block and the physical--low block both exhibit an identical density (34\%). This confirms that high- and low-skilled occupations do not differ in their utilization of specialized skills within their respective domains. The more salient difference lies in the distribution of general skills. The general--high block is nearly saturated (density of 94\%), whereas the general--low block is only about half filled (density of 59\%). This discrepancy implies that the skilled-level of an occupation may depend not only on specialization, but also on how fluently it can interact with others.

While workers in the labor market reap returns from their productive activities utilizing specialized competencies, each individual faces inherent limits in the quantity of capabilities they can possess \cite{neffke2019value}. The homogeneity of occupational diversity captures this constraint (Fig.~\ref{fig:Figure S5}). Overcoming such individual limitations has historically relied on coordination and exchange their competencies with other workers. The results of this section can be interpreted through the lens of core concepts from various disciplines. In terms of ecology, generalist species mitigate inter-species competition, thereby enhancing the survivability of specialist species \cite{mariani2019physicsreport}. In the context of human capital, recent research has emphasized the role of general skills---particularly social and interactive skills---in reducing coordination costs and facilitating task trade across jobs \cite{deming2017growing}. In this vein, the role of general skills can be interpreted through the lens of economic complexity: general skills reduce the cost of forming links within the network, thereby facilitating the greater accumulation of knowledge and know-how \cite{hidalgo2015information}.

\subsection*{Wage Premium of General Skills}

\begin{figure}[H]
    \centering
    \includegraphics[width=\textwidth]{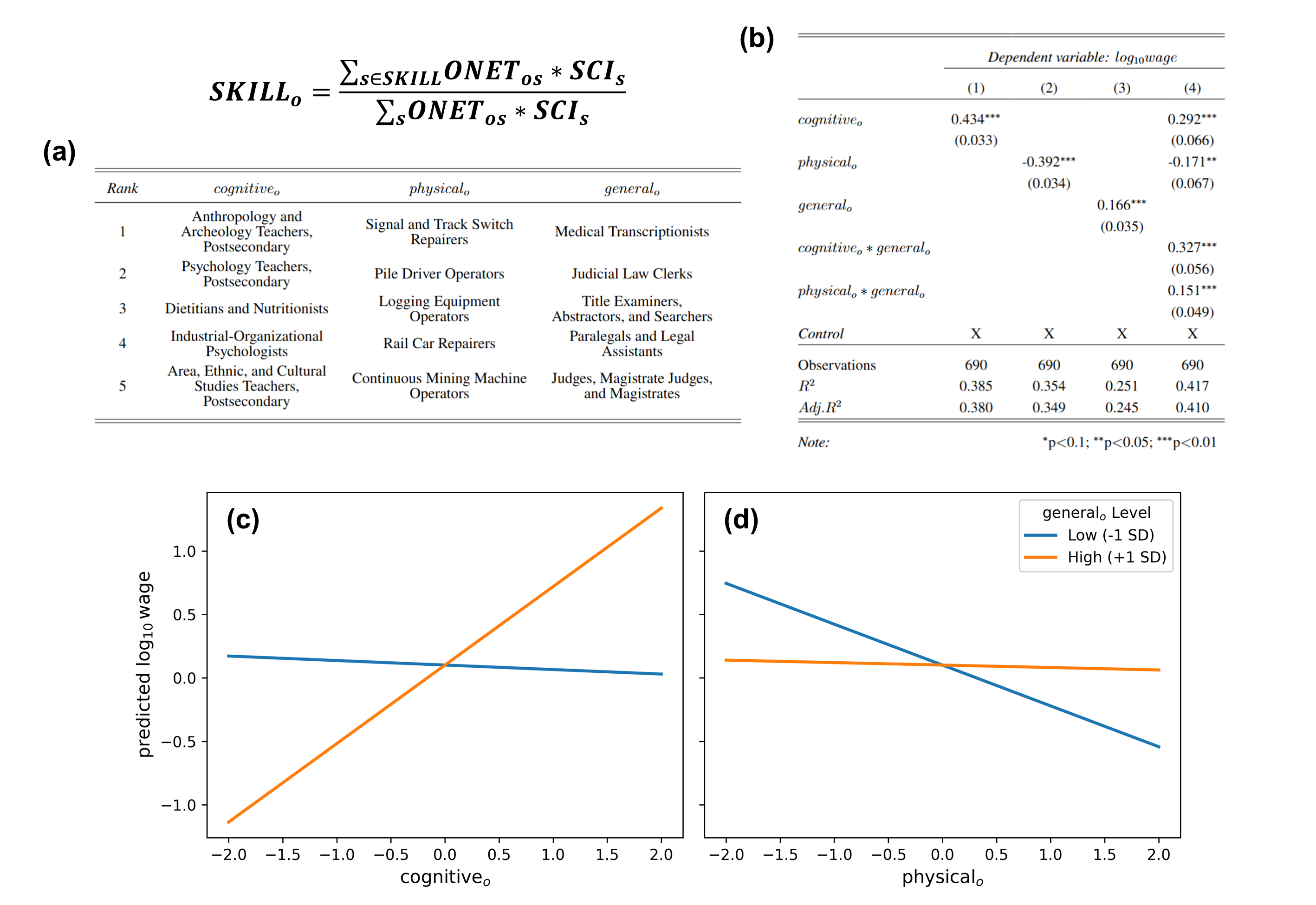}
    \caption{\textbf{Wage Effects of Skill Composition and the Role of General Skills.} \textbf{(a)} Top five occupations by skill type based on the highest values of $SKILL_o$. $SKILL$ indicates each skill community (cognitive, physical, and general). $ONET_{os}$ is the importance of each skill for a given occupation from the O*NET data. $SCI_s$ is min-max normalized to [0, 1]. \textbf{(b)} OLS regression results with interaction terms between $general_o$ and the two specialized skill types. Following Alabdulkareem et al. \cite{alabdulkareem2018unpacking}, we control for the share of employment within each occupation by required level of education: less than a high school diploma, more than a high school diploma, more than a bachelor’s degree, and more than a graduate school diploma. In Model (4), $general_o$ is omitted due to a linear dependency among the three skill categories (i.e., their weights sum to one). Thus, the coefficients for $cognitive_o$ and $physical_o$ should be interpreted relative to the baseline of general skills. All data are aggregated at the 6-digit SOC code and subsequently standardized. \textbf{(c)} Predicted wage by $cognitive_o$ for low ($-1$ SD) and high ($+1$ SD) levels of $general_o$, showing that general skills amplify the wage premium of cognitive skills. \textbf{(d)} Predicted wage by $physical_o$ for different levels of $general_o$, indicating that general skills mitigate the wage penalty of physical skills.}
    \label{fig:Figure 4}
\end{figure}

To examine the role of general skills in the labor market outcome, we estimate the wage premiums of each skill type as well as the interaction effects between general and specialized skills. The independent variables ($SKILL_o$) are constructed by weighting the importance of each skill for a given occupation ($ONET_{os}$) by its $SCI_s$. For each occupation $o$, the weighted values are then aggregated within each skill-type category (i.e., cognitive, physical, and general), and normalized by the total weighted sum across all skills. This yields the relative contribution of each skill type to an occupation’s overall skill composition which is captured by the OCI. 

Figure~\ref{fig:Figure 4} (a) presents the top five occupations that require the highest proportion of each skill type. Occupations with the highest reliance on cognitive skills are predominantly high-OCI occupations including postsecondary teaching positions. In contrast, those that rely most heavily on physical skills include low-skilled occupations such as \textit{Signal and Track Switch Repairers}, \textit{Pile Driver Operators}, and \textit{Logging Equipment Operators}. Occupations that demand general skills the most include middle-skilled jobs such as \textit{Medical Transcriptionists}, \textit{Judicial Law Clerks}, and \textit{Title Examiners, Abstractors, and Searchers}.

Figure~\ref{fig:Figure 4} (b) illustrates how the effects of cognitive and physical skills on occupational wages are moderated by general skills. As a first step, we investigate the wage effects of each skill type. Model (1) shows that cognitive skills are positively associated with wages, indicating that occupations with a higher reliance on cognitive skills which are embedded in upper hierarchy in the labor market structure tend to offer higher wages. In contrast, Model (2) reveals a strong negative association between physical skills and wages, confirming that occupations primarily requiring physical skills---which are structurally embedded in the lower tiers of the labor market---tend to be less rewarded. Model (3) incorporates general skills, which are positively related to wages. However, when compared to Model (1) and (2), the specification with only $general_o$ exhibits a lower explanatory power ($R^2 = 0.251$), showing that general skills have a comparatively weaker direct effect on wages.

Model (4) introduces interaction terms between general skills and each of the two specialized skill types to assess whether the returns to specialized skills vary with the level of general skills. In the most restrictive specification (Model 4), the significant interaction effect between $cognitive_o$ and $general_o$ ($0.327$, $p < 0.01$) is larger than the main effect of $cognitive_o$ ($0.292$, $p < 0.01$), indicating that general skills significantly amplify the wage returns to cognitive skills. This finding is consistent with recent evidence on the complementarity between social and cognitive skills in the labor market \cite{deming2017growing}.

For physical skills, general skills exhibit an even more pronounced moderating effect. In the full specification, the main effect of $physical_o$ remains negative and statistically significant ($-0.171$, $p < 0.05$), confirming that physical skill-intensive occupations are penalized in the labor market when considered in isolation. However, the interaction term between $physical_o$ and $general_o$ is positive and significant ($0.151$, $p < 0.01$), indicating that the wage penalty associated with physical skills diminishes and even reverses in occupations where general skills are also highly required. In other words, the marginal effect of physical skills on wages changes sign depending on the level of general skills. This striking reversal highlights a strong complementary relationship, whereby general skills not only offset the disadvantages of physical skills but potentially convert them into wage advantages. These results underscore the structural role of general skills in enhancing the labor market value of both cognitive and physical skills.

Figure~\ref{fig:Figure 4} (c) and (d) provide a visual illustration of the interaction effects described above. Figure~\ref{fig:Figure 4} (c) shows that the positive association between $cognitive_o$ and wages is substantially steeper when $general_o$ is high, reflecting that general skills amplify the wage returns to cognitive skills. Figure~\ref{fig:Figure 4} (d) reveals that the negative relationship between $physical_o$ and wages weakens and flattens at higher levels of $general_o$, highlighting the compensatory role of general skills in offsetting the wage penalty associated with physical skills.

In sum, the analysis demonstrates that general skills play a critical moderating role in shaping the wage effects of specialized skills. Structurally embedded at the core of the occupation--skill network, general skills enable workers to translate their specialized competencies into wage gains by facilitating broader exchange opportunities in the labor market. They amplify the wage premium of cognitive skills and mitigate the wage penalty of physical skills. These findings suggest that the value of specialized skills depends on the presence of general skills, which supports favorable wage outcomes across diverse occupations.

\section*{Discussion and Conclusion}

The division of labor has driven prosperity in the modern economy by enabling specialization based on individual competencies and facilitating their exchange, thereby generating synergies among workers \cite{neffke2019value,hosseinioun2025skill}. Human capacity is inherently limited; no individual can possess the full range of skills. Therefore, the only way to overcome this constraint is to coordinate with others across occupations through general skills, thereby avoiding isolation in the labor market.

This study explores the structural embeddedness of skills in the labor market. Our findings are threefold. First, general skills, which are required across many occupations, are distinct from the two types of specialized skills: cognitive and physical. They primarily encompass social skills that are essential for interactions within the labor market. Second, we uncover the embeddedness of skills within the labor market structure. Applying the Method of Reflections to the occupation--skill bipartite network, we derive complexity indices that compress connectivity in the labor market. Furthermore, we unpack the structure of the labor market with these indices: general skills are centrally embedded, whereas cognitive and physical skills are positioned at the upper and lower ends of the hierarchy, forming two independent nested structures with general skills. Lastly, we estimate the moderating effect of general skills on the returns to specialized skills. General skills amplify the positive wage effects of cognitive skills and mitigate---even reverse---the wage penalties associated with physical skills.

The findings from current study contribute to the literature that examines human capital and labor market structure through network analysis. While traditional human capital theory distinguishes between general and specialized (or specific) skills, existing network studies often emphasize the latter. By leveraging the skill importance data from O*NET, which already has normalized information, we bridge the gap between human capital theory and the network-based approach, revealing the central embeddedness of general skills in the labor market structure. Furthermore, while previous studies have primarily utilized complexity indices as measures of economic sophistication \cite{balland2019smart,balland2017rigby,chun2024hur,pinheiro2022unrelated}, our analysis reveals the modular and nested structural properties of the labor market that arises from occupational skill requirements.

The implications of this study extend beyond academia to practical applications. Based on the main findings from the current study, reskilling strategies should move beyond highlighting occupation-specific skills. The existing papers often suggest that low-skilled workers must abruptly transform their skill sets to overcome disadvantages in the labor market. However, such approaches entail substantial social and individual burdens and may prove unrealistic. A more feasible path lies in developing general skills as a foundation for the reintegration of low-skilled workers, thereby overcoming isolation from the labor market.

Nevertheless, this study faces several limitations. First, although the Method of Reflections is a powerful tool for deriving complexity indices, its results can be sensitive to binarization thresholds. Second, recent studies have increasingly used job posting data, which provide more granular information on  skills than the O*NET database \cite{stephany2024price,anderson2017skill,bornerevans2018skillPNAS,deming2020noray,tong2021evans}. While such granularity allows for more precise observation of labor dynamics, it also entails a trade-off with the curse of dimensionality. Future research could provide a more dynamic understanding of the labor market by integrating the methodology proposed in the current study with more detailed information extracted from job postings.

\section*{Data and Methods}

\textbf{O*NET and OEWS data.} The U.S. Bureau of Labor Statistics (BLS) provides O*NET (Occupational Information Network) and OEWS (Occupational Employment and Wage Statistics) data. O*NET contains occupational characteristics, including importance of skills, education level requirements, and work context of occupations. Because of the finer-grained information about occupations in the U.S., a large body of studies analyzed the O*NET database. The 120 skills in this study correspond to the O*NET variables of Knowledge, Skills, and Abilities (KSA), follows \cite{hosseinioun2025skill}. 

O*NET investigates occupational data by its own classification system, 8-digit O*NET-SOC codes, which is a more granular system than Standard Occupational Classification. Employing O*NET 28.0 (Aug. 2023), we examine 872 O*NET-SOC occupations and, when presenting the results, aggregate them by averaging at a higher-level classification as appropriate. Specifically, O*NET relies on OEWS data, which maps several information to occupations including annual wage and employment of each occupation, at the 6-digit Detailed Classification. In this situations, we summarize the values of occupations to the 6-digit occupational classifications to construct regression data. OEWS data in May 2023 is utilized.

\vspace{\baselineskip}
\noindent\textbf{Constructing the Skill Network.} Following the convention in the economic complexity literature, we binarize the data to make the skill requirements of occupations more explicit. We take into account that the skill importance values from the O*NET database are measured on a Likert scale. RCA normalization employed in previous studies is typically applied to values such as dollars or employment, which can take on an unbounded range. For these reasons, we adopt a simpler approach by applying a threshold to the skill importance values when binarizing the data. 

The resulting matrix \textbf{M} is a bipartite matrix in which rows correspond to occupations and columns to skills. If $M_{os} = 1$, it indicates that the skill $s$ is required for occupation $o$; otherwise, it is 0. To construct the skill network, we compute the pairwise proximity between skills, following the equation introduced by \cite{alabdulkareem2018unpacking,hidalgo2007product}:
\[
proximity_{ss'} = \frac{\sum_{o} M_{os} M_{os'}}{\max\left( \sum_{o} M_{os}, \sum_{o} M_{os'} \right)}
\]
It reflects how often two skills are required for the same occupations. This equation yields the weights of 7,140 edges between 120 skill nodes.

To intuitively classify the skills, we apply the Louvain algorithm to the skill network. The Louvain community detection algorithm is a widely used heuristic method in network studies. We repeat the algorithm 100 times on the skill network, and three stable communities are extracted. 

\vspace{\baselineskip}
\noindent\textbf{Method of Reflections.}
Hidalgo and Hausmann \cite{hidalgo2009building} introduced the Method of Reflections (MoR) to derive the complexity of countries and products through the connectivity of nodes in the country--product bipartite network. The MoR is an iterative algorithm that compresses the network’s structural connectivity into a one-dimensional representation. At each iteration, the value of a node is updated by averaging the values of its directly connected nodes from the previous order:
\[
k_{o,n}= \frac{\sum_{s} M_{os}k_{s,n-1}}{k_{o,0}}
\]
\[
k_{s,n}= \frac{\sum_{o} M_{os}k_{o,n-1}}{k_{s,0}}
\]
Here, $k_{o,0}= \sum_{s} M_{os}$ and $k_{s,0}= \sum_{o} M_{os}$ represent the degree centrality of occupational and skill nodes in the occupation--skill network, respectively. We label $k_{o,0}$ as the diversity of an occupation, reflecting the number of skills it requires. Similarly, $k_{s,0}$, referred to as the ubiquity of a skill, indicates the number of occupations that require a given skill.

As shown in Figure~\ref{fig:Figure 2}, the values of each node gradually converge, capturing not only direct but also mediated connectivity. For an occupation, the first-order value reflects the average ubiquity of its required skills, while for a skill, it represents the average diversity of the occupations that require it. The second-order values are more nuanced: for occupations, they reflect the average diversity of other occupations sharing the same skill set; for skills, they indicate the average ubiquity of skills that co-occur within the same occupations \cite{hidalgo2009building}.

As the iteration order increases, the value assigned to an occupation (or skill) increasingly compresses information not only about its directly connected skills (or occupations), but also about those indirectly connected through shared nodes. Consequently, occupations (or skill) with similar connectivity profiles converge to similar values, whereas those with distinct structures diverge. For instance, although cognitive and physical skills have similar ubiquity, they diverge in higher orders due to being connected to distinct sets of occupations (Fig.~\ref{fig:Figure 2}).

At sufficiently high orders, the values of nodes converge and compress the structural connectivity in the bipartite network. These converged values correspond to the eigenvector associated with the second largest eigenvalue of the network matrix, which can also be obtained through spectral clustering \cite{mealy2019interpreting}. This contrasts slightly with approaches that use the Fitness algorithm---an alternative measure in economic complexity---to estimate the complexity of occupations and skills, as in Aufiero et al. \cite{aufiero2024mapping}. We define the standardized high-order values of occupations and skills as the Occupational Complexity Index (OCI) and the Skill Complexity Index (SCI), respectively.

\vspace{\baselineskip}
\noindent\textbf{The Contribution of Skills to Modularity and Nestedness}
The literature on economic complexity and ecology share a common foundation in network science. In ecology, modularity or nestedness have been widely used to characterize the structural properties of bipartite networks, such as those formed by species interactions (e.g., plant--pollinator networks) \cite{mariani2019physicsreport}. In contrast, the economic complexity literature has, to our knowledge, paid little attention to such structural properties, excepting \cite{bustos2012nestedPlosOne,ren2020nestedHSSC,miao2022YY}. To overcome this limitation, we apply metrics from the ecology literature to examine the structure of the occupation--skill network, which exhibits both modular and nested properties.

Modularity of the occupation--skill network is defined as follows \cite{barber2007modularity}:
\[
Q = \frac{1}{m} \sum_{o} \sum_{s} (M_{os} - P_{os}) \, \delta(community_o, community_s)
\]
$m$ is the total number of connections in \textbf{M}. $P_{os}$ denotes the probability that a connection between occupation $o$ and skill $s$ exists in \textbf{M}. This probability is derived from the null model, where $P_{os} = \frac{diversity_o \cdot ubiquity_s}{m}$. $\delta$ equals 1 if $community_o = community_s$, and 0 otherwise. Based on the observation in Figure~\ref{fig:Figure 3} (c), we assume that cognitive skills--high-skilled occupations, general skills--middle-skilled occupations, and physical skills--low-skilled occupations each belong to the same community, respectively.

Among various metrics developed to measure the nestedness of a bipartite matrix, NODF (Nestedness metric based on Overlap and Decreasing Fill) is the most widely used due to its capacity to account for both row- and column-wise overlap. However, calculating the NODF can be computationally intensive for large matrices. To effectively represent the nestedness and to align with the objective of the current study, which focuses on understanding the nestedness structure of the occupation--skill network arising from the embeddedness of skills, we employ a simplified measurement of nestedness based solely on skill-wise overlap, which nevertheless produces results closely aligned with those of NODF \cite{hosseinioun2025skill}. The overlap degree $N$ is defined as follows \cite{wrightreeves1992Nc_nest}:
\[
N = \sum_s \frac{ubiquity_s(ubiquity_s - 1)}{2}
\]
This expression calculates the number of possible co-occurrence pairs for each skill across occupations and sums them over all skills, thereby capturing the extent to which skills are commonly required together across occupations.

Hosseinioun et al. \cite{hosseinioun2025skill} introduce a null model to estimate the contribution of individual skills to the nestedness of the occupation--skill network, building on the approach proposed by Saavedra et al. \cite{saavedra2011uzzi_nested}. We extend this method to evaluate the contribution of each skill to both modularity ($C_s^Q$) and nestedness ($C_s^N$) of the network. The contribution scores are defined as follows:
\[C_s^Q=(Q - <Q_s^*>)/\sigma_{Q_s^*}\]
\[C_s^N=(N- <N_s^*>)/\sigma_{N_s^*}\]
where $Q_s^*$ and $N_s^*$ represent the modularity and nestedness of the network computed from randomized simulations in which only the connections of skill $s$ are reshuffled, preserving the ubiquity of a given skill. For each skill, we perform 1,000 simulations and compute the corresponding mean and standard deviation. A positive value of $C_s$ indicates that skill $s$ contributes significantly more than expected by chance to the structural properties of the network, whereas a negative value suggests the opposite.

\section*{Acknowledgements}

This research was supported by the National Science Foundation Grant Award Number
EF–2133863.

\section*{Author Contributions}
SL, DJ, and JL conceived the study. DJ and JL supervised and coordinated the project. SL and DJ performed the analyses presented in the article. SL and DJ contributed to writing the manuscript. All authors reviewed and edited the manuscript, approved the final version for publication, and agreed to be held accountable for the work presented herein.

\bibliographystyle{unsrt}
\bibliography{references}

\clearpage
\renewcommand{\thefigure}{S\arabic{figure}}
\setcounter{figure}{0}

\maketitle
\section*{Supplementary Information}

\begin{figure}[H]
    \centering
    \includegraphics[width=\textwidth]{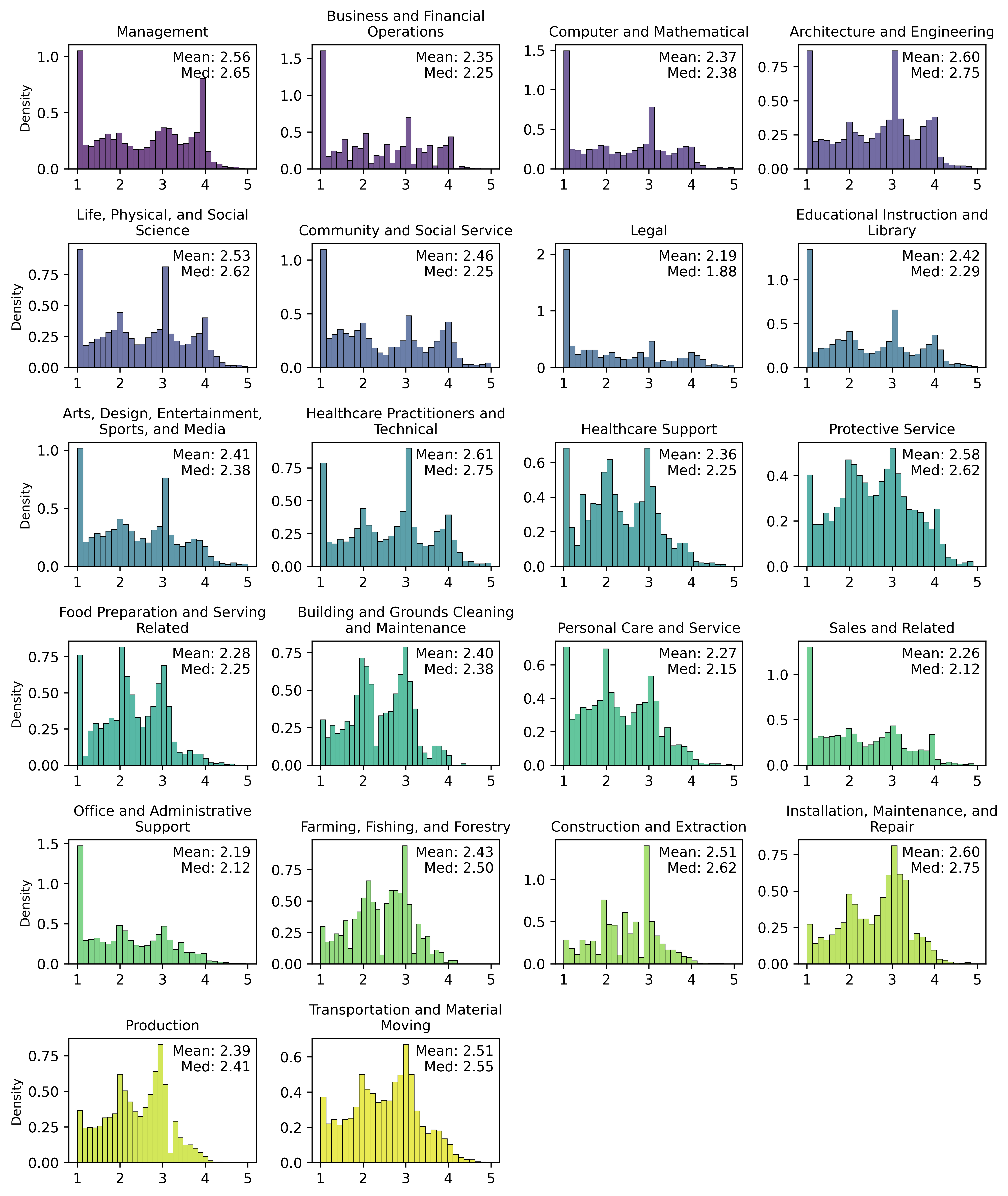}
    \caption{\textbf{Distribution of Skill Importance Values by 2-digit Major Occupations.} }
    \label{fig:Figure S1}
\end{figure}

\begin{figure}[H]
    \centering
    \includegraphics[width=\textwidth]{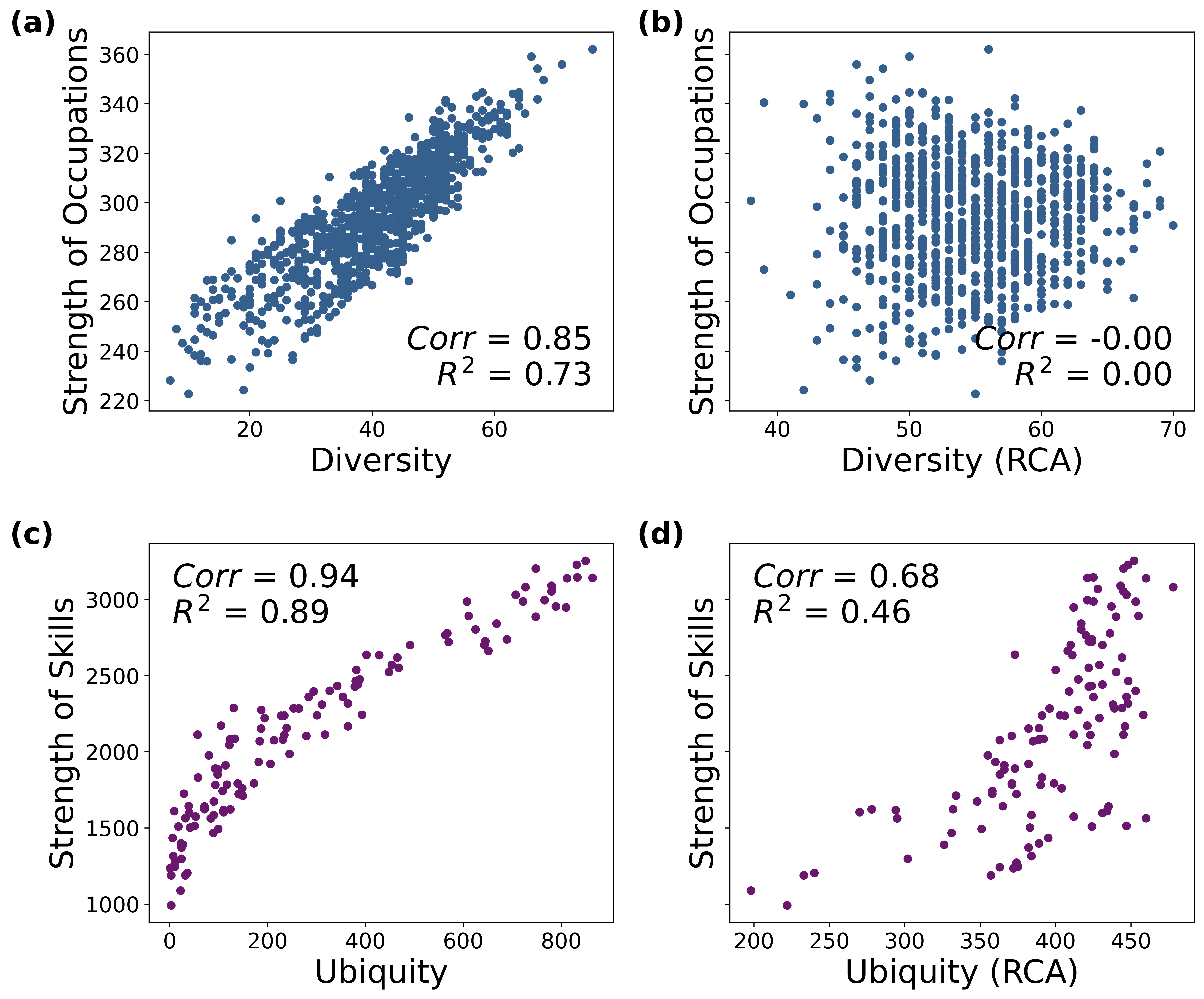}
    \caption{\textbf{The relationship between node degrees in \textbf{M} and node strengths in the original data.} The plots compare the correlation between node degrees (diversity and ubiquity) and strength (sum of skill importance) in the binary occupation--skill matrix \textbf{M} versus RCA-based networks. Strong correlations in the binarized case indicate that the simplified matrix preserves information from the original data.}
    \label{fig:Figure S2}
\end{figure}

\begin{figure}[H]
    \centering
    \includegraphics[width=\textwidth]{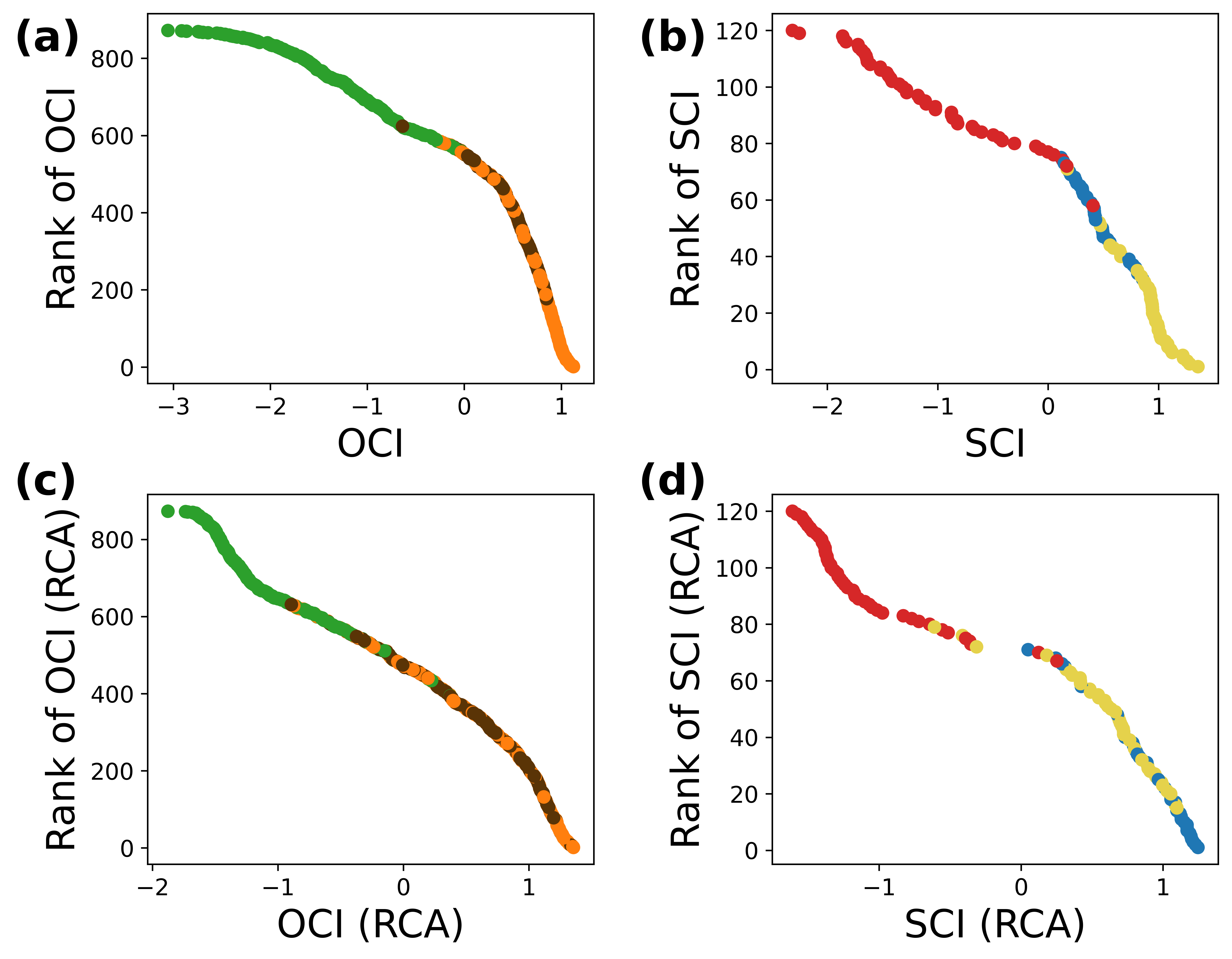}
    \caption{\textbf{The Relationship between Complexity Indices' Values and Rankings.} Despite capturing structural patterns, the continuous nature of complexity indices prevents them from identifying distinct clusters \cite{mealy2019interpreting}. However, the SCI derived using RCA, which highlights over-expressed skills, reveals a clear separation of skills.}
    \label{fig:Figure S3}
\end{figure}

\begin{figure}[H]
    \centering
    \includegraphics[width=0.5\textwidth]{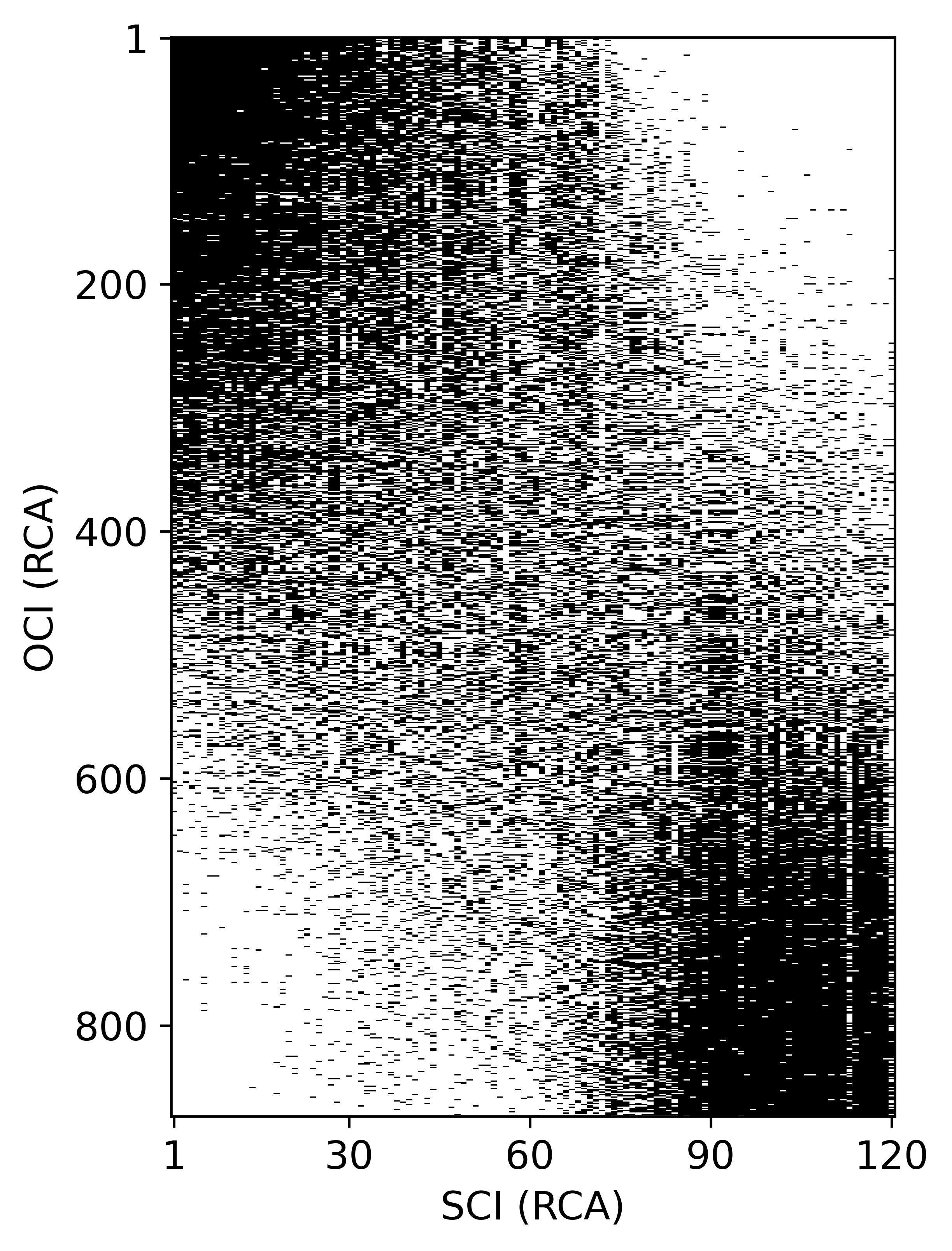}
    \caption{\textbf{Heatmap of M$^{RCA}$ arranged by the rankings of OCI$^{RCA}$ and SCI$^{RCA}$} reveals a modular pattern, but not a nested one.}
    \label{fig:Figure S4}
\end{figure}

\begin{figure}[H]
    \centering
    \includegraphics[width=\textwidth]{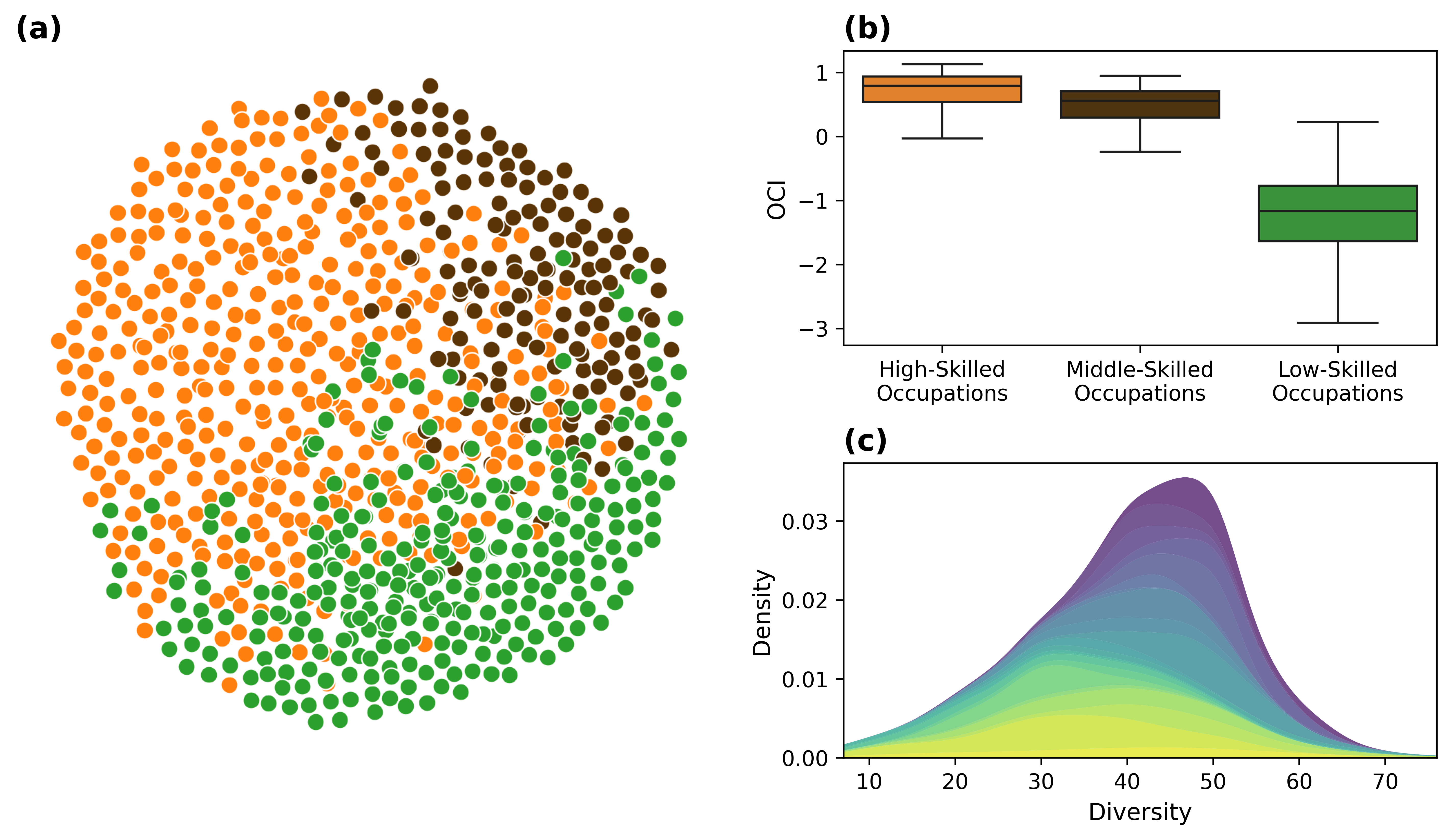}
    \caption{\textbf{Occupation Network and the Distributions of Diversity and OCI.} \textbf{(a)} The occupation network consisting of 872 occupations. Similar to community detection in the skill network, we applied the Louvain algorithm 100 times to the occupation network and identified three stable communities. Based on the distribution of OCI values within each community, we labeled them as High-, Middle-, and Low-Skilled Occupations. \textbf{(b)} The distribution of OCI values in each occupational community. \textbf{(c)} The distribution of occupational diversity is approximately normal, suggesting that the human capacity to carry skills is relatively homogeneous and bounded. Stacked colors correspond to the 2-digit Major Occupation groups.}
    \label{fig:Figure S5}
\end{figure}

\end{document}